\documentclass[aps,prb,twocolumn,amssymb,amsmath]{revtex4}
\usepackage{amsmath}
\usepackage{amssymb}
\usepackage{graphicx}

\begin{document}

\title{Critical fluctuation conductivity in layered superconductors in
strong
electric field}

\author{I. Puica}

\altaffiliation[Also at ]{Department of Physics, Polytechnic University
of Bucharest, Spl. Independentei 313, RO-77206 Bucharest 6, Romania}

\email{ipuica@ap.univie.ac.at}

\author{W. Lang}

\affiliation{Institut f\"{u}r Materialphysik der Universit\"{a}t Wien,
Boltzmanngasse 5, A-1090 Wien, Austria}

\begin{abstract}
The paraconductivity, originating from critical superconducting
order-parameter fluctuations in the vicinity of the critical
temperature in a layered superconductor is calculated in the frame of
the self-consistent Hartree approximation, for an arbitrarily strong
electric field and zero magnetic field. The paraconductivity diverges
less steep towards the critical temperature in the Hartree
approximation than in the Gaussian one and it shows a distinctly
enhanced variation with the electric field. Our results indicate that
high electric fields can be effectively used to suppress
order-parameter fluctuations in high-temperature superconductors.
\end{abstract}

 \maketitle

\section{Introduction}

Due to their high critical temperature, small coherence length, and
quasi-two-dimensional nature, the high-temperature superconductors
(HTSC) show a much more pronounced and therefore experimentally
accessible effect of thermodynamic fluctuations in the critical region
of the normal-superconducting transition. In general, an enhancement of
the conductivity, denoted paraconductivity, is observed in HTSC above
$T_{c}$ due to the presence of superconducting fluctuations. Outside
the critical region, in the absence of the magnetic field and for small
electric fields, the paraconductivity can be explained in terms of the
Aslamazov-Larkin\cite{Aslamazov68} theory of noninteracting, Gaussian
fluctuations. The initial expressions for the paraconductivity have
been extended for two-dimensional layered superconductors, a situation
very much resembling the crystal structure in the cuprates, by Lawrence
and Doniach.\cite{Lawrence71} However, it was shown that the
fluctuation conductivity may be calculated in the linear-response
approximation only for sufficiently weak fields, when they do not
perturb the fluctuation spectrum.\cite{Hurault69} Reasonably high
values of the electric field $E$ can accelerate the fluctuating paired
carriers to the depairing current, and thus, suppress the lifetime of
the fluctuations, leading to deviation from Ohm's law. In connection
with the low-temperature superconductors, the nonlinearity has been
studied theoretically for the isotropic case\cite{Schmid69,Tsuzuki70}
and also proven experimentally on thin aluminum films.\cite{Thomas71}
The issue of the non-ohmic fluctuation conductivity for a clean layered
superconductor in an arbitrary electric field has been addressed by
Varlamov and Reggiani,\cite{Varlamov92} starting from a microscopic
approach of Gor'kov\cite{Gorkov70} for dirty isotropic superconductors.
Essentially the same dependence on temperature and electric field has
been recently\cite{Mishonov02} derived, together with generalizations
for the case of arbitrary dimension, based on the analytical derivation
of the velocity distribution resulting from the Boltzmann equation for
the fluctuating Cooper pairs.

The above-mentioned theories do not consider the interactions between
fluctuations, so that fluctuations can be described by the Gaussian
approximation. Thus, the quartic term in the Ginzburg-Landau free
energy is neglected. This approximation is known to hold for
temperature values not too close to the mean-field transition
temperature, but it breaks down in the critical region, since the
nonlinear character of the TDGL equation cannot be neglected for high
densities of fluctuation Cooper pairs. Several
works\cite{Hassing73,Quader88,Fisher91,Ullah91,Dorsey91,Ikeda91a,Ikeda91b,Nishio97,Wickham0}
have included the interaction between superconducting fluctuations in
the critical transition region within different theoretical approaches.
The simplest and most used one is the Hartree approximation which
treats self-consistently the quartic term in the Ginzburg-Landau
free-energy expansion. In this way expressions for the specific heat
have been derived for bulk\cite{Hassing73} and layered\cite{Quader88}
superconductors under magnetic field, based on the functional integral
approach. In the frame of the time-dependent Ginzburg-Landau (TDGL)
theory, Ullah and Dorsey\cite{Ullah91} computed the Nernst effect, the
thermopower, the longitudinal and the Hall conductivity in the
linear-response approximation for a layered superconductor in a
magnetic field. Using the same relaxational dynamics of the TDGL
approach, Dorsey\cite{Dorsey91} provided expressions for the
fluctuation conductivity in both ohmic and non-ohmic regime, for
isotropic superconductors of arbitrary dimensionality and in the
absence of a magnetic field. More recently \cite{Wickham0} the effects
of critical superconducting fluctuations on the scaling of the linear
ac conductivity, $\sigma (\omega )$, of a bulk superconductor slightly
above $T_{c}$ in zero applied magnetic field have been investigated
based on the dynamic renormalization-group method applied to the
relaxational TDGL model of superconductivity, verifying explicitly the
scaling hypothesis $\sigma (\omega ,\xi )$ proposed originally by
Fisher, Fisher, and Huse.\cite{Fisher91} The essential features of the
scaling and renormalization-group method has been also reviewed
recently by Larkin and Varlamov.\cite{Larkin02} A more sophisticated
approach based on renormalization procedure and diagrammatic techniques
has been developed for treating the non-Gaussian superconducting
fluctuations by Ikeda, Ohmi and Tsuneto,\cite{Ikeda91a} and applied to
the longitudinal conductivity, the magnetization\cite{Ikeda91b} and the
Hall conductivity\cite{Nishio97} in the linear-response approximation,
for a layered superconductor under magnetic field.

In this paper we shall address the problem of the non-ohmic behavior of
the non-Gaussian fluctuation conductivity for a layered superconductor,
a topic that, to our present knowledge, has not yet been treated in the
literature. While the non-linear conductivity under arbitrarily strong
electric field was derived for a layered
system\cite{Varlamov92,Mishonov02} for Gaussian, noninteracting
fluctuations, the effect of the critical, strongly interacting
fluctuation on the ohmic and non-ohmic conductivity was
investigated\cite{Dorsey91} only for isotropic systems of arbitrary
dimensionality, but not for the layered Lawrence-Doniach model. The
latter, however, would be required for comparison to experimental data
on, e.g., YBa$_{2}$Cu$_{3}$O$_{6+x}$. The paper is organized as
follows. In Section \ref{Model-eq} the TDGL equations are deduced for a
layered superconducting system, with the explicit consideration of an
arbitrarily strong electric field, oriented parallel to the layers. The
fluctuation interaction term is also included in this model, within the
self-consistent Hartree approximation. Section \ref{UVcutoff} presents
the resulting equation solutions, obtained with the aid of the Green
function technique (detailed in the Appendix). By considering in detail
the necessary correction through the UV cut-off procedure, expressions
for the fluctuation conductivity and the self-consistent equation for
the renormalized reduced temperature parameter are provided. Further,
Section \ref{LimitCases} treats the limit cases of the linear response
approximation, the no-cut-off limit, and also the isotropic 2D and 3D
cases, recovering thus results of previous theories. In Section
\ref{Results} the application of the model is illustrated by comparing
the paraconductivity obtained in the present theory with that in the
Gaussian fluctuation approximation for various applied electric fields.
A comparison between the fluctuation suppression effects of the
electric and magnetic fields is also illustrated. Finally, in Section
\ref{Conclusion}, we summarize the main conclusions emerging from our
analysis.

\section{TDGL equation for arbitrary electric field}

\label{Model-eq}For our purpose, we shall adopt the TDGL framework, and
treat the quartic term in the free-energy expansion within the simple
self-consistent Hartree approximation. The starting point will be the
Lawrence-Doniach expression of the Ginzburg-Landau (GL) free energy for
a system of superconducting planes separated by a distance $s$, with a
Josephson coupling between the planes, in the absence of magnetic
field, \begin{eqnarray}
\mathfrak{F}=\sum _{n}\int d^{2}x\left[a\left|\psi
_{n}\right|^{2}+\frac{\hbar ^{2}}{2m}\left|\nabla \psi
_{n}\right|^{2}\right. &  & \nonumber \\
\left.+\frac{\hbar ^{2}}{2m_{c}s^{2}}\left|\psi _{n}-\psi
_{n+1}\right|^{2}+\frac{b}{2}\left|\psi _{n}\right|^{4}\right]\; , &  &
\label{FreeEn}
\end{eqnarray}
 where $m$ and $m_{c}$ are effective Cooper pair masses in the
$ab$-plane
and along the $c$-axis, respectively. The GL potential
$a=a_{0}\varepsilon $ is parameterized by $a_{0}=\hbar ^{2}/2m\xi
_{0}^{2}$ and $\varepsilon =\ln \left(T/T_{0}\right)\approx
\left(T-T_{0}\right)/T_{0}$, with $T_{0}$ being the mean-field
transition temperature and $\xi _{0}$ the in-plane GL-coherence length,
extrapolated at $T=0$. The critical dynamics of the complex
superconducting order parameter $\psi _{n}$ in the $n$-th plane will be
described by the gauge-invariant relaxational time-dependent
Ginzburg-Landau equation \begin{equation} \Gamma
_{0}^{-1}\left(\frac{\partial }{\partial t}+i\frac{e_{0}}{\hbar
}\varphi \right)\psi _{n}=-\frac{\delta \mathfrak{F}}{\delta \psi
_{n}^{*}}+\zeta _{n}\left(\mathbf{x},t\right)\;
,\label{TDGLeq}\end{equation}
 where the pair electric charge is $e_{0}=2e$, and the order parameter
relaxation rate $\Gamma _{0}$, given
by\cite{Masker69,Cyrot73,Ovchinnikov01}\begin{equation} \Gamma
_{0}^{-1}=\frac{\pi \hbar ^{3}}{16m\xi _{0}^{2}k_{B}T}\:
,\label{Gamma0}\end{equation}
 is related to the life-time of metastable Cooper
pairs\cite{Larkin02,Mishonov02}
$\tau ^{(\mathrm{BCS})}\approx \pi \hbar /16k_{B}(T-T_{0})$ through the
relation $\Gamma _{0}^{-1}=2a\tau ^{(\mathrm{BCS})}$. As it can be
noticed, we define the relaxation rate $\Gamma _{0}^{-1}$ as depending
on the actual temperature $T$, while in many other
works,\cite{Schmid66,Hurault69,Larkin02} it is defined as a function of
the mean-field critical temperature, $T_{0}$. Of course the difference
is negligible near the transition point, but one must recall that in
the derivation of the GL equations from the BCS
theory,\cite{Gorkov59,Gennes66,Cyrot73} the temperature is usually
approximated with the critical one, while the reduced temperature $\ln
\left(T/T_{0}\right)$ is approximated with
$\left(T-T_{0}\right)/T_{0}$. The same happens while the TDGL equation
is obtained for temperatures near $T_{0}$, as for instance in Refs.
\onlinecite{Schmid66} and \onlinecite{Abrahams66}. In some more recent
derivations of the TDGL,\cite{Krempaski85,Ovchinnikov01} however, the
original appearance of these parameters is preserved, so that the order
parameter relaxation rate is written as depending on the actual
temperature. In this mathematically somewhat stricter sense, one could
therefore write the life-time of the metastable Cooper pairs as $\tau
^{(\mathrm{BCS})}=\pi \hbar /16k_{B}T\ln \left(T/T_{0}\right)$.

The Langevin forces $\zeta _{n}\left(\mathbf{x},t\right)$ introduced in
Eq. (\ref{TDGLeq}) in order to model the thermodynamical fluctuations
must satisfy the fluctuation-dissipation theorem, and ensure that the
system relaxes to the proper equilibrium distribution. This requirement
is fulfilled if the Langevin forces $\zeta
_{n}\left(\mathbf{x},t\right)$ are correlated by the Gaussian
white-noise law \begin{equation} \left\langle \zeta
_{n}\left(\mathbf{x},t\right)\zeta
_{n'}^{*}\left(\mathbf{x}',t'\right)\right\rangle =2\Gamma
_{0}^{-1}k_{B}T\delta (\mathbf{x}-\mathbf{x}')\delta (t-t')\frac{\delta
_{nn'}}{s}\; ,\label{Noise}\end{equation}
 where $\delta (\mathbf{x}-\mathbf{x}')$ is the 2-dimensional
delta-function
concerning the in-plane coordinates. Since we are interested in finding
the conductivity for an arbitrary electric field, we cannot use the
linear-response approximation, so we have to explicitly include the
electric field in the model. In order to compute the in-plane
fluctuation conductivity, we shall assume the field $\mathbf{E}$ along
the $x$-axis (where $x$ and $y$ are the in-plane coordinates),
generated by the scalar potential $\varphi =-Ex$. In the chosen gauge,
the current density operator along the $x$ direction in the $n$-th
plane will be given by \begin{equation} j_{x}^{(n)}=-\frac{ie_{0}\hbar
}{2m}\left[\psi _{n}^{*}(\mathbf{x},t)\partial _{x}\psi
_{n}(\mathbf{x},t)-\psi _{n}(\mathbf{x},t)\partial _{x}\psi
_{n}^{*}(\mathbf{x},t)\right]\; ,\label{Current}\end{equation}
 so that after averaging with respect to the noise, \begin{equation}
\left\langle j_{x}^{(n)}\right\rangle =\left.-\frac{ie_{0}\hbar
}{2m}(\partial _{x}-\partial _{x'})\left\langle \psi
_{n}\left(\mathbf{x},t\right)\psi
_{n}^{*}\left(\mathbf{x}',t\right)\right\rangle
\right|_{\mathbf{x}=\mathbf{x}'}\; .\label{CurrentDef}\end{equation}
 The time-dependent GL equation (\ref{TDGLeq}) writes \begin{eqnarray}
\Gamma _{0}^{-1}\frac{\partial \psi _{n}}{\partial
t}-i\frac{e_{0}\Gamma _{0}^{-1}Ex}{\hbar }\psi _{n}-\frac{\hbar
^{2}}{2m}\nabla ^{2}\psi _{n} &  & \nonumber \\
+a\psi _{n}+\frac{\hbar ^{2}}{2m_{c}s^{2}}(2\psi _{n}-\psi _{n+1}-\psi
_{n-1}) &  & \nonumber \\
+b\left|\psi _{n}\right|^{2}\psi _{n}=\zeta
_{n}\left(\mathbf{x},t\right)\; . &  & \label{EQini}
\end{eqnarray}
 As already mentioned, the quartic term in the thermodynamical
potential
will be treated in the Hartree approximation, in the same sense as
applied also in previous works,\cite{Masker69,Ullah91,Penev0} namely by
replacing the cubic term $b\left|\psi _{n}\right|^{2}\psi _{n}$ in Eq.
(\ref{EQini}) with $b\left\langle \left|\psi
_{n}\right|^{2}\right\rangle \psi _{n}$. In this way, the non-linearity
is decoupled, resulting in a linear problem with a modified
(renormalized) GL potential $\widetilde{a}=a+b\left\langle \left|\psi
_{n}\right|^{2}\right\rangle $, which implies a renormalized reduced
temperature \begin{equation} \widetilde{\varepsilon }=\varepsilon
+\frac{b}{a_{0}}\left\langle \left|\psi _{n}\right|^{2}\right\rangle \;
.\label{RenormEps}\end{equation}
 The average $\left\langle \left|\psi _{n}\right|^{2}\right\rangle $
is to be determined, in principle, self-consistently together with the
parameter $\widetilde{\varepsilon }$.

In order to simplify the following computations, we shall introduce the
Fourier transform with respect to the two in-plane coordinates and the
layer index, respectively, through the relations:\begin{eqnarray}
\psi _{n}(\mathbf{x},t) & = & \int \frac{d^{2}\mathbf{k}}{(2\pi
)^{2}}\int _{-\pi /s}^{\pi /s}\frac{dq}{2\pi }\psi
_{q}(\mathbf{k},t)e^{-i\mathbf{xk}}e^{-iqns}\nonumber \\
\psi _{q}(\mathbf{k},t) & = & \int d^{2}\mathbf{x}\sum _{n}s\, \psi
_{n}(\mathbf{x},t)e^{i\mathbf{xk}}e^{iqns}\; ,\label{Fourier}
\end{eqnarray}
 so that Eq. (\ref{EQini}) becomes \begin{eqnarray}
\left[\Gamma _{0}^{-1}\frac{\partial }{\partial t}-\frac{e_{0}\Gamma
_{0}^{-1}E}{\hbar }\frac{\partial }{\partial _{k_{x}}}+\frac{\hbar
^{2}\mathbf{k}^{2}}{2m}+\widetilde{a}\right. &  & \label{EQFourier}\\
\left.+\frac{\hbar ^{2}\gamma ^{2}}{ms^{2}}(1-\cos qs)\right]\psi
_{q}(\mathbf{k},t) & = & \zeta _{q}\left(\mathbf{k},t\right)\;
.\nonumber
\end{eqnarray}
 We have introduced the anisotropy parameter $\gamma =\xi _{0c}/\xi
_{0}=\sqrt{m/m_{c}}$,
with $\xi _{0c}$ the out-of-plane GL-coherence length extrapolated at
$T=0$. The term $\zeta _{q}\left(\mathbf{k},t\right)$ in Eq.
(\ref{EQFourier}) is the Fourier transform according to the rules
(\ref{Fourier}) of the noise function $\zeta
_{n}\left(\mathbf{x},t\right)$. One can directly verify that the
following correlation function holds:
\begin{eqnarray}
\left\langle \zeta _{q}\left(\mathbf{k},t\right)\zeta
_{q'}^{*}\left(\mathbf{k}',t'\right)\right\rangle  & = & 2\Gamma
_{0}^{-1}k_{B}T(2\pi )^{3}\label{FourNoiseCorr}\\
 &  & \cdot \delta (\mathbf{k}-\mathbf{k}')\delta (q-q')\delta (t-t')\,
.\nonumber
\end{eqnarray}
 Equation (\ref{EQFourier}) may be solved using the Green function
method, as done also by Tucker and Halperin,\cite{Tucker71} and
Dorsey\cite{Dorsey91} in order to solve the relaxational TDGL equation
with a Langevin noise for isotropic superconductors. Our derivation
differs however by the fact that it is applied for a layered
superconductor, and also through the differently chosen potential
gauge. We denote thus with $R_{q}(\mathbf{k},t;k'_{x},t')$ the Green
function for the Eq. (\ref{EQFourier}), which satisfies
\begin{eqnarray}
\left[\Gamma _{0}^{-1}\frac{\partial }{\partial t}-\frac{e_{0}\Gamma
_{0}^{-1}E}{\hbar }\frac{\partial }{\partial _{k_{x}}}+\frac{\hbar
^{2}k_{x}^{2}}{2m}+a_{1}\right]R_{q}(\mathbf{k,}t;k'_{x},t') &  &
\nonumber \\
=\delta (k_{x}-k'_{x})\delta (t-t')\; , &  & \label{GreenEq}
\end{eqnarray}
 where we have introduced the notation \begin{equation}
a_{1}\equiv \widetilde{a}+\frac{\hbar ^{2}k_{y}^{2}}{2m}+\frac{\hbar
^{2}\gamma ^{2}}{ms^{2}}(1-\cos qs)\; .\label{NotatA1}\end{equation}
 Thus, the solution of Eq. (\ref{EQFourier}) will then be written
\begin{equation}
\psi _{q}(\mathbf{k},t)=\int dt'\int dk'_{x}\;
R_{q}(\mathbf{k,}t;k'_{x},t')\zeta _{q}\left(k'_{x},k_{y},t'\right)\;
.\label{Convol}\end{equation}

The Green function $R_{q}(\mathbf{k,}t;k'_{x},t')$ is computed in the
Appendix, and given by Eq. (\ref{GreenSol}), so that the relation
(\ref{Convol}) for the Fourier transformed order parameter will write
accordingly\begin{widetext}\begin{eqnarray}
\psi _{q}(\mathbf{k},t) & = & \Gamma _{0}\exp \left\{ \frac{\hbar
\Gamma _{0}}{e_{0}E}\left[\frac{\hbar
^{2}k_{x}^{3}}{6m}+a_{1}k_{x}\right]\right\} \int dt'\; \theta
(t-t')\label{ConvolSol}\\
 &  & \cdot \exp \left\{ -\frac{\hbar \Gamma
_{0}}{e_{0}E}\left[\frac{\hbar ^{2}\left(k_{x}+\frac{e_{0}E}{\hbar
}\left[t-t'\right]\right)^{3}}{6m}+a_{1}\left(k_{x}+\frac{e_{0}E}{\hbar
}\left[t-t'\right]\right)\right]\right\} \nonumber \\
 &  & \cdot \zeta _{q}\left(k_{x}+\frac{e_{0}E}{\hbar
}\left[t-t'\right],\; k_{y},\; t'\right)\, .\nonumber
\end{eqnarray}
 One can notice that the solution (\ref{ConvolSol}) fulfills
causality,
due to the retarded character of the Green function
$R_{q}(\mathbf{k,}t;k'_{x},t')$. It can be also simplified further,
\begin{eqnarray} \psi _{q}(\mathbf{k},t) & = & \Gamma _{0}\int
_{0}^{\infty }d\tau \exp \left\{ -\Gamma _{0}\left[a_{1}\tau
+\frac{\tau \hbar ^{2}}{2m}\left(k_{x}+\frac{e_{0}E\tau }{2\hbar
}\right)^{2}+\frac{e_{0}^{2}E^{2}}{24m}\tau ^{3}\right]\right\} \zeta
_{q}\left(k_{x}+\frac{e_{0}E\tau }{\hbar },\; k_{y},\; t-\tau \right)\,
.\label{IntegrTau1}
\end{eqnarray}
 Now, following (\ref{CurrentDef}), the current density can be written
as\begin{eqnarray}
\left\langle j_{x}^{(n)}\right\rangle  & = & -\frac{e_{0}\hbar
}{m}2\Gamma _{0}k_{B}T\int \frac{dk_{x}}{2\pi }\int \frac{dk_{y}}{2\pi
}\int _{-\pi /s}^{\pi /s}\frac{dq}{2\pi }\: k_{x}\cdot
\label{CurrDensGen}\\
 &  & \cdot \int _{0}^{\infty }d\tau \, \exp \left\{ -\Gamma
_{0}\left[2\left(\widetilde{a}+\frac{\hbar
^{2}k_{y}^{2}}{2m}+\frac{\hbar ^{2}\gamma ^{2}}{ms^{2}}\left[1-\cos
qs\right]\right)\tau +\frac{\tau \hbar
^{2}}{m}\left(k_{x}+\frac{e_{0}E\tau }{2\hbar
}\right)^{2}+\frac{e_{0}^{2}E^{2}}{12m}\tau ^{3}\right]\right\} \,
,\nonumber
\end{eqnarray}
 while the averaged density of fluctuating Coooper pairs writes
analogously\begin{eqnarray}
 &  & \left\langle \left|\psi _{n}\right|^{2}\right\rangle =2\Gamma
_{0}k_{B}T\int \frac{dk_{x}}{2\pi }\int \frac{dk_{y}}{2\pi }\int _{-\pi
/s}^{\pi /s}\frac{dq}{2\pi }\label{CooperDensGen}\\
 &  & \cdot \int _{0}^{\infty }d\tau \, \exp \left\{ -\Gamma
_{0}\left[2\left(\widetilde{a}+\frac{\hbar
^{2}k_{y}^{2}}{2m}+\frac{\hbar ^{2}\gamma ^{2}}{ms^{2}}\left[1-\cos
qs\right]\right)\tau +\frac{\tau \hbar
^{2}}{m}\left(k_{x}+\frac{e_{0}E\tau }{2\hbar
}\right)^{2}+\frac{e_{0}^{2}E^{2}}{12m}\tau ^{3}\right]\right\} \,
,\nonumber
\end{eqnarray}
\end{widetext}where we have taken into account the expression
(\ref{FourNoiseCorr})
for the Fourier transformed noise correlation function, and also
replaced the notation $a_{1}$ with its value (\ref{NotatA1}).

Before proceeding further and solve the integrals over momentum
variables in Eqs. (\ref{CurrDensGen}) and (\ref{CooperDensGen}), we
must recall the inherent ultraviolet (UV) divergence of the
Ginzburg-Landau theory, which is not valid on length scales less than
the zero-temperature coherence length $\xi _{0}$. The short wavelength
fluctuations break down the {}``slow variation condition'' for the
superconducting order parameter, a central hypothesis of the GL
approach.\cite{TinkhamBook} This difficulty can be solved by applying
an UV cut-off to the fluctuation spectrum, a procedure introduced from
the beginning in the GL theory.\cite{LandauBook}

\section{Solutions with consideration of the UV cut-off}

\label{UVcutoff}The classical\cite{Johnson76,Schmid69,Penev0} procedure
is to suppress the short wavelength fluctuating modes through the
\emph{momentum cut-off} condition \begin{equation} \mathbf{k}^{2}<c\xi
_{0}^{-2}\; ,\label{Cut-off1}\end{equation}
 where the dimensionless cut-off factor $c$ is close to unity. Also
a \emph{total energy cut-off} was
suggested,\cite{Patton69,Nam72,Carballeira01} which eliminates the most
energetic fluctuations and not only those with short wavelengths,
\begin{equation} \mathbf{k}^{2}+\xi ^{-2}(\varepsilon )<c\xi
_{0}^{-2}\: .\label{Cut-off2}\end{equation}
 Heuristically, the adequacy of the total-energy cut-off can be
justified
on the grounds of the Gaussian GL approach by taking into account that
the probability of each fluctuating mode is controlled by its total
energy $\hbar ^{2}\mathbf{k}^{2}/2m+a_{0}\varepsilon $, and not only by
its momentum.\cite{Larkin02,Mishonov02} Very recently\cite{Vidal02} it
was suggested that the physical meaning of the {}``total energy''
cut-off follows from the uncertainty principle, which impose a limit to
the confinement of the superconducting wave function. It must be
however mentioned that in the critical fluctuation region, the two
cut-off prescriptions almost coincide quantitatively, due to the low
reduced-temperature value $\varepsilon $ with respect to the factor
$c$.

For simplicity, in the following we shall apply the cut-off procedure
in its classical form (\ref{Cut-off1}) on the $\mathbf{k}$-plane
momentum integrals in Eqs. (\ref{CurrDensGen}) and
(\ref{CooperDensGen}). If necessary, one can get easily also the result
that would correspond to the {}``total energy'' cut-off
(\ref{Cut-off2}), by simply replacing $c\rightarrow
c-\widetilde{\varepsilon }$. Although for the Lawrence-Doniach model,
the $z$-axis momentum $q$ also contributes to the fluctuation mode
energy with the term $\left(\hbar ^{2}\gamma
^{2}/ms^{2}\right)\left(1-\cos qs\right)$, the inclusion of a momentum
cut-off in this direction is not necessary, since the $z$-axis spectrum
is already modulated through $-\pi /s\leq q\leq \pi /s$. One can thus
perform the integral over $q$-momentum in Eqs. (\ref{CurrDensGen}) and
(\ref{CooperDensGen}) and get\begin{eqnarray}
\int _{-\pi /s}^{\pi /s}\frac{dq}{2\pi }\exp \left[-2\Gamma _{0}\tau
\frac{\hbar ^{2}\gamma ^{2}}{ms^{2}}\left(1-\cos qs\right)\right] &  &
\nonumber \\
=\frac{1}{s}\exp \left(-\frac{2\Gamma _{0}\hbar ^{2}\gamma ^{2}\tau
}{ms^{2}}\right)I_{0}\left(\frac{2\Gamma _{0}\hbar ^{2}\gamma ^{2}\tau
}{ms^{2}}\right)\, , &  & \label{IntegrQ}
\end{eqnarray}
 where we used the identity $I_{0}(x)=(1/2\pi )\int _{-\pi }^{\pi
}e^{x\cos \theta }d\theta $
for the modified Bessel function $I_{0}(x)$.

We shall therefore apply the cut-off by performing the momentum
integral from Eq. (\ref{CurrDensGen}) for $\mathbf{k}^{2}<c\xi
_{0}^{-2}$,\begin{widetext}\begin{eqnarray}
 &  & \int \limits _{\mathbf{k}^{2}<c\xi
_{0}^{-2}}\frac{d^{2}\mathbf{k}}{\left(2\pi \right)^{2}}\, k_{x}\exp
\left\{ -2\Gamma _{0}\tau \frac{\hbar
^{2}}{2m}\left[k_{y}^{2}+\left(k_{x}+\frac{e_{0}E\tau }{2\hbar
}\right)^{2}\right]\right\} \label{Cut-off-Current}\\
 &  & =e^{-\frac{\Gamma _{0}e_{0}^{2}E^{2}\tau
^{3}}{4m}}\frac{1}{\left(2\pi \right)^{2}}\int _{0}^{\sqrt{c}\xi
_{0}^{-1}}dk\, k^{2}\, e^{-\frac{\tau \Gamma _{0}\hbar
^{2}k^{2}}{m}}\int _{-\pi }^{\pi }d\varphi \, \cos \varphi \,
e^{-\frac{\Gamma _{0}\tau ^{2}\hbar e_{0}Ek\cos \varphi }{m}}\nonumber
\\
 &  & =-e^{-\frac{\Gamma _{0}e_{0}^{2}E^{2}\tau ^{3}}{4m}}\frac{1}{2\pi
}\left(\frac{ma_{0}}{\hbar ^{2}}\right)^{3/2}\int _{0}^{c}dw\,
\sqrt{2w}\, e^{-2\tau \Gamma _{0}a_{0}w}\, I_{1}\left(\frac{\Gamma
_{0}\tau ^{2}e_{0}E}{m}\sqrt{2ma_{0}w}\right)\, ,\nonumber
\end{eqnarray}
\end{widetext}where we introduced the new dimensionless variable
$w=\hbar ^{2}k^{2}/2ma_{0}$ and used the first order modified Bessel
function $I_{1}(x)=(1/2\pi )\int _{-\pi }^{\pi }d\varphi \, \cos
\varphi \, e^{x\cos \varphi }$. The current density (\ref{CurrDensGen})
will write eventually, after considering Eqs. (\ref{IntegrQ}) and
(\ref{Cut-off-Current}), and introducing the new integration variable
$u=2a_{0}\Gamma _{0}\tau $,
\begin{eqnarray}
 &  & j(\widetilde{\varepsilon },E)=\frac{ek_{B}T}{\pi \hbar s\xi
_{0}}\int _{0}^{\infty }du\, I_{0}\left(\frac{2\gamma ^{2}\xi
_{0}^{2}}{s^{2}}u\right)e^{-\left(\widetilde{\varepsilon
}+\frac{2\gamma ^{2}\xi _{0}^{2}}{s^{2}}\right)u}\label{Current-cut}\\
 &  & \cdot e^{-4\left(\frac{\pi e\xi
_{0}E}{16\sqrt{3}k_{B}T}\right)^{2}u^{3}}\int _{0}^{c}dw\,
\sqrt{w}e^{-uw}I_{1}\left(\frac{\pi e\xi
_{0}}{8k_{B}T}Eu^{2}\sqrt{w}\right)\; ,\nonumber
\end{eqnarray}
 where we have also expressed the parameters $\Gamma _{0}$ and $a_{0}$
with the aid of the in-plane coherence length $\xi _{0}$.

In order to infer the fluctuation conductivity $\sigma =j/E$, we shall
use the fact that the function $I_{1}(x)$ consists only of odd argument
powers, and satisfies namely the identity
$I_{1}(x)=(x/2)\left[I_{0}(x)-I_{2}(x)\right]$, so that we can finally
write the paraconductivity under the
form\begin{widetext}\begin{eqnarray}
\sigma (\widetilde{\varepsilon },E) & = & \frac{e^{2}}{16\hbar s}\int
_{0}^{\infty }du\, I_{0}\left(\frac{ru}{2}\right)\,
u^{2}e^{-\left(\widetilde{\varepsilon
}+\frac{r}{2}\right)u-4\left(\frac{E}{E_{0}}\right)^{2}u^{3}}\label{Sigma-cutted-off}\\
 &  & \cdot \int _{0}^{c}dw\, w\,
e^{-uw}\left[I_{0}\left(2\sqrt{3}\frac{E}{E_{0}}
u^{2}\sqrt{w}\right)-I_{2}\left(2\sqrt{3}\frac{E}{E_{0}}u^{2}
\sqrt{w}\right)\right]\, .\nonumber
\end{eqnarray}
\end{widetext}where we have introduced the notations\begin{equation}
r=\frac{4\gamma ^{2}\xi _{0}^{2}}{s^{2}}=\left(\frac{2\xi
_{0c}}{s}\right)^{2}\quad \mathrm{and}\quad
E_{0}=\frac{16\sqrt{3}k_{B}T}{\pi e\xi
_{0}}\label{e-E0-def}\end{equation}
 for the anisotropy parameter $r$, and the characteristic electric
field $E_{0}$, respectively. Relations (\ref{e-E0-def}) can be
expressed also as depending on microscopical parameters, like the Fermi
velocity $v_{F}$ and the electronic interlayer hopping energy $J$. By
identifying the in-plane GL-coherence length $\xi _{0}$ from the
microscopic derivation of the GL equation in the two-dimensional case,
one has in the clean limit\cite{Larkin02} $\xi _{0}=\left(7\zeta
(3)/32\right)^{1/2}\hbar \, v_{F}/\pi k_{B}T_{0}$, which implies for
$T\approx T_{0}$\begin{equation} E_{0}=64\sqrt{\frac{6}{7\zeta
(3)}}\frac{k_{B}^{2}T_{0}^{2}}{e\hbar v_{F}}\quad \mathrm{and}\quad
r=\frac{7\zeta (3)}{8\pi ^{2}}\frac{J^{2}}{k_{B}^{2}T_{0}^{2}}\,
,\label{E0Varl-rDorin}\end{equation}
 where we have also used in the same approximations (clean limit and
$T\approx T_{0}$), the expression for the anisotropy parameter $r$ from
Ref. \onlinecite{Dorin93}, as a function of the interlayer electron
hopping energy $J$.

In an analogous manner as presented above for the current-density, one
can apply the cut-off procedure on the momentum integral in Eq.
(\ref{CooperDensGen}), and obtain

\begin{eqnarray}
 &  & \int \limits _{\mathbf{k}^{2}<c\xi
_{0}^{-2}}\frac{d^{2}\mathbf{k}}{\left(2\pi \right)^{2}}\exp \left\{
-2\Gamma _{0}\tau \frac{\hbar
^{2}}{2m}\left[k_{y}^{2}+\left(k_{x}+\frac{e_{0}E\tau }{2\hbar
}\right)^{2}\right]\right\} \label{Cut-off-Compute}\\
 &  & =e^{-\frac{\Gamma _{0}e_{0}^{2}E^{2}\tau
^{3}}{4m}}\frac{ma_{0}}{2\pi \hbar ^{2}}\int _{0}^{c}dw\, e^{-2\tau
\Gamma _{0}a_{0}w}I_{0}\left(\frac{\Gamma _{0}\tau
^{2}e_{0}E}{m}\sqrt{2ma_{0}w}\right)\, ,\nonumber
\end{eqnarray}
 so that the averaged density of Cooper pairs (\ref{CooperDensGen})
becomes\begin{widetext}\begin{equation} \left\langle \left|\psi
_{n}\left(\mathbf{x},t\right)\right|^{2}\right\rangle
=\frac{mk_{B}T}{2\pi \hbar ^{2}s}\int _{0}^{\infty }du\,
I_{0}\left(\frac{ru}{2}\right)\, e^{-\left(\widetilde{\varepsilon
}+\frac{r}{2}\right)\, u-4\left(\frac{E}{E_{0}}\right)^{2}u^{3}}\int
_{0}^{c}dw\,
e^{-uw}I_{0}\left(2\sqrt{3}\frac{E}{E_{0}}u^{2}\sqrt{w}\right)\,
,\label{Mod2tris}\end{equation}
 where we passed from $\tau $ to the variable $u$, and used the
notations (\ref{e-E0-def}). One can easily prove that without the UV
cut-off (i.e. for $c\rightarrow \infty $), the expression would be
divergent. Replacing $c\rightarrow c-\widetilde{\varepsilon }$ would in
turn correspond to the energy cut-off under the form (\ref{Cut-off2}).
The self-consistent equation (\ref{RenormEps}) for the parameter
$\widetilde{\varepsilon }$ will write therefore:\begin{equation}
\widetilde{\varepsilon }=\ln \frac{T}{T_{0}}+gT\int _{0}^{\infty }du\,
I_{0}\left(\frac{ru}{2}\right)\, e^{-\left(\widetilde{\varepsilon
}+\frac{r}{2}\right)\, u-4\left(\frac{E}{E_{0}}\right)^{2}u^{3}}\int
_{0}^{c}dw\,
e^{-uw}I_{0}\left(2\sqrt{3}\frac{E}{E_{0}}u^{2}\sqrt{w}\right)\;
,\label{EpsTilde}\end{equation}
\end{widetext}where the factor \begin{equation}
g=\frac{2\mu _{0}\kappa ^{2}e^{2}\xi _{0}^{2}k_{B}}{\pi \hbar
^{2}s}\label{factor}\end{equation}
 has been computed by taking into account the expression of the
quartic
term coefficient\cite{TinkhamBook} $b=\mu _{0}\kappa ^{2}e_{0}^{2}\hbar
^{2}/2m^{2}$, with $\kappa $ being the Ginzburg-Landau parameter
$\kappa =\lambda _{0}/\xi _{0}$.

The above Eq. (\ref{Sigma-cutted-off}) for the paraconductivity,
together with the self-consistent Eq. (\ref{EpsTilde}), both valid for
an arbitrary strong electric field and with the explicit inclusion of
the UV cut-off, are the main results of this paper. A short comment
would be useful regarding the application of the cut-off procedure to
the $\mathbf{k}$-integrals in Eqs. (\ref{Cut-off-Compute}) and
(\ref{Cut-off-Current}). It could seem more appealing to apply the
cut-off on the translated wave-vector magnitude, such as
$k_{y}^{2}+\left[k_{x}+\left(e_{0}E\tau /2\hbar \right)\right]^{2}<c\xi
_{0}^{-2}$, as it was approximately done, although in a different
gauge, by Kajimura and Mikoshiba,\cite{Kajimura71} while studying the
paraconductivity in arbitrary electric field in the two-dimensional
case. This would certainly simplify the calculations and consequently
the factors that depend on the cut-off parameter $c$, but would not
really correspond to the actual meaning of the UV cut-off, which is to
assure the slow variation condition for the order parameter by
eliminating the rapidly oscillating modes in the Fourier expansion
(\ref{Fourier}). It can be seen for instance that when the dummy
variable $\tau $ grows towards $\infty $, the component $k_{x}$ would
have to approach $-\infty $ in order to preserve the cut-off condition
in the translated form. It can be moreover verified that the
{}``translated'' cut-off would not give the correct result in the $E=0$
limit, namely the third term in the Eq. (\ref{LD-cutOff}) below would
be missing.

\section{Limit cases\label{LimitCases}}

\subsection{Linear response limit}

It is worth comparing to previous results what become Eqs.
(\ref{Sigma-cutted-off}) and (\ref{EpsTilde}) in the zero-field limit,
$E=0$. Taking into account that $I_{0}(0)=1$ and $I_{2}(0)=0$, and
using the integral form for the modified Bessel function
$I_{0}\left(ru/2\right)$, one obtains for the linear response
conductivity \begin{eqnarray}
\left.\sigma (\widetilde{\varepsilon })\right|_{E=0} & = &
\frac{e^{2}}{16\hbar s}\int _{0}^{\infty }du\; \left(1-e^{-c\, u}-c\,
u\, e^{-c\, u}\right)\nonumber \\
 &  & \cdot I_{0}\left(\frac{r}{2}u\right)\exp
\left(-\widetilde{\varepsilon }\, u-\frac{r}{2}u\right)\nonumber \\
 & = & \frac{e^{2}}{16\hbar
s}\left[\frac{1}{\sqrt{\widetilde{\varepsilon
}\left(\widetilde{\varepsilon
}+r\right)}}-\frac{1}{\sqrt{\left(\widetilde{\varepsilon
}+c\right)\left(\widetilde{\varepsilon }+c+r\right)}}\right.\nonumber
\\
 &  & \left.-\frac{c(c+\widetilde{\varepsilon
}+r/2)}{\left[(c+\widetilde{\varepsilon }+r)(c+\widetilde{\varepsilon
})\right]^{3/2}}\right]\; ,\label{LD-cutOff}
\end{eqnarray}
 which is the Lawrence-Doniach\cite{Lawrence71} formula for the
fluctuation
conductivity of a layered superconductor, with the inclusion of the UV
cut-off. A formally identical expression was also inferred by
Carballeira \emph{et al.},\cite{Carballeira01} for Gaussian
fluctuations (i.e. with $\widetilde{\varepsilon }=\varepsilon $), in
order to fit the HTSC paraconductivity in the high reduced-temperature
region.

The integral in Eq. (\ref{EpsTilde}) can also be easily performed in
the limit $E=0$, and it yields the relation \begin{eqnarray}
\widetilde{\varepsilon }-\ln \frac{T}{T_{0}} & = & \frac{2\mu
_{0}\kappa ^{2}e^{2}\xi _{0}^{2}k_{B}T}{\pi \hbar ^{2}s}\int
_{0}^{c}\frac{dx}{\sqrt{\left(x+\widetilde{\varepsilon
}\right)\left(x+\widetilde{\varepsilon }+r\right)}}\nonumber \\
 & = & \frac{4\mu _{0}\kappa ^{2}e^{2}\xi _{0}^{2}k_{B}T}{\pi \hbar
^{2}s}\ln \frac{\sqrt{\widetilde{\varepsilon
}+c}+\sqrt{\widetilde{\varepsilon }+c+r}}{\sqrt{\widetilde{\varepsilon
}}+\sqrt{\widetilde{\varepsilon }+r}}\; ,\label{AnalogUD}
\end{eqnarray}
that matches the formula found by Mishonov and Penev,\cite{Penev0} with
the only difference that in Ref. \onlinecite{Penev0} the temperature
$T$ is approximated with $T_{0}$ in the logarithm prefactor. It can be
easily verified that Eq. (\ref{AnalogUD}) is also coincident with the
analogous equation found previously by Ullah and Dorsey\cite{Ullah91}
(UD) for the renormalized reduced temperature $\widetilde{\varepsilon
}_{H}$, if one considers the zero magnetic field limit in the UD
formula. In this limit, the sum over Landau levels transforms, with the
aid of Euler-MacLaurin summation formula, to the integral from Eq.
(\ref{AnalogUD}), if one takes also into account that the cut-off used
by UD corresponds with $c=2$.

We have thus shown that from our Eqs. (\ref{Sigma-cutted-off}) and
(\ref{EpsTilde}) one can infer in the limit $E\rightarrow 0$ the
already known results for conductivity and renormalized reduced
temperature in the ohmic approximation.

\subsection{No-cut-off limit}

The cut-off procedure is crucial for calculating the averaged
fluctuating Cooper pairs density, since Eq. (\ref{CooperDensGen})
yields a divergent $\tau $-integral when one performs the
$\mathbf{k}$-momentum integrals on the entire $\mathbf{k}$-plane.

The result (\ref{Sigma-cutted-off}) for the paraconductivity remains
however finite even if one removes the cut-off (i.e. for $c\rightarrow
\infty $), although it will give then a larger paraconductivity than in
the cut-off case, especially for higher reduced temperatures
($\varepsilon \geq 0.1$). Turning back to Eq. (\ref{CurrDensGen}) with
the cut-off removed, one is allowed to translate the integral variable
$k_{x}$ so that $k_{x}+\left(e_{0}E\tau /2\hbar \right)\rightarrow
k_{x}$, and after performing the Poisson $\mathbf{k}$-integrals, one
obtains eventually for the fluctuation in-plane conductivity in the
presence of an arbitrary electric field $E$, without considering the UV
cut-off, the simpler relation \begin{equation} \sigma
(\widetilde{\varepsilon },E)=\frac{e^{2}}{16\hbar s}\int _{0}^{\infty
}du\, I_{0}\left(\frac{r}{2}u\right)\cdot e^{-\widetilde{\varepsilon
}\, u-\frac{r}{2}u-\left(\frac{E}{E_{0}}\right)^{2}u^{3}}\,
.\label{Sigma}\end{equation}

The expression (\ref{Sigma}) is not new. In its form, it is essentially
similar to the ones found by Varlamov and Reggiani\cite{Varlamov92} and
Mishonov \emph{et al.}\cite{Mishonov02} for the case of Gaussian
fluctuations, if one neglects the difference which consists in the
presence of the renormalized parameter $\widetilde{\varepsilon }$ in
Eq. (\ref{Sigma}) instead of the reduced temperature $\varepsilon =\ln
(T/T_{0})$. It must be stated however that Ref.
\onlinecite{Varlamov92}, based on a microscopical
approach,\cite{Gorkov70} defines an out-of-plane coherence length
larger by a factor $\sqrt{2}$ than the commonly used one. Nevertheless,
its result corresponds to our Eq. (\ref{Sigma}) when expressed through
Eqs. (\ref{E0Varl-rDorin}) in the microscopical parameters $v_{F}$ and
$J$ (the latter is denoted in Ref. \onlinecite{Varlamov92} by $w$).
Mishonov \emph{et al.}\cite{Mishonov02} signalize differences in one of
their intermediary results with respect to Refs. \onlinecite{Gorkov70}
and \onlinecite{Varlamov92}, but they find the same dependence of the
non-ohmic conductivity on the reduced temperature $\varepsilon $ and
the electric field $E$, by solving the Boltzmann equation for the
velocity distribution of the fluctuating Cooper pairs. Their formula
differs mathematically from Eq. (\ref{Sigma}) only by the presence of
two extra-factors before the integral, namely the ratio between $a$-
and $b$- coherence lengths (since Ref. \onlinecite{Mishonov02} takes
into account also a general in-plane anisotropy), and the ratio
$T/T_{0}$, which in fact comes artificially only if one writes the
relaxation rate $\Gamma _{0}$ with $T_{0}$ instead of $T$ (see the
comment after Eq. \ref{Gamma0}).

\subsection{Isotropic limit}

Results analogous to Eqs. (\ref{Sigma-cutted-off}) and (\ref{EpsTilde})
for the isotropic two dimensional (2D) case can be easily derived by
simply taking the limit $\xi _{0c}=0$ (or $r\rightarrow 0$) and
identifying the interlayer distance $s$ with the film thickness $d$. We
will have thus\begin{widetext}\begin{eqnarray}
\sigma ^{(2\mathrm{D})}(\widetilde{\varepsilon },E) & = &
\frac{e^{2}}{16\hbar d}\int _{0}^{\infty }du\,
u^{2}e^{-\widetilde{\varepsilon }\,
u-4\left(\frac{E}{E_{0}}\right)^{2}u^{3}}\int _{0}^{c}dw\, w\,
e^{-uw}\left[I_{0}\left(2\sqrt{3}\frac{E}{E_{0}}u^{2}
\sqrt{w}\right)-I_{2}\left(2\sqrt{3}\frac{E}{E_{0}}u^{2}\sqrt{w}\right)\right]\,
,\label{Sigma-2D}\\
\widetilde{\varepsilon }-\ln \frac{T}{T_{0}} & = & \frac{2\mu
_{0}\kappa ^{2}e^{2}\xi _{0}^{2}k_{B}T}{\pi \hbar ^{2}d}\int
_{0}^{\infty }du\, e^{-\widetilde{\varepsilon }\,
u-4\left(\frac{E}{E_{0}}\right)^{2}u^{3}}\int _{0}^{c}dw\,
e^{-uw}I_{0}\left(2\sqrt{3}\frac{E}{E_{0}}u^{2}\sqrt{w}\right)\,
.\label{EpsTilde2D}
\end{eqnarray}
\end{widetext}

Equations (\ref{Sigma-2D}) and (\ref{EpsTilde2D}) differ from the
analogous results of Ref. \onlinecite{Kajimura71}, because there only
an approximate form of the cut-off procedure was applied, as we have
already mentioned at the end of Section \ref{UVcutoff}.

If one neglects the cut-off ($c\rightarrow \infty $), Eq.
(\ref{EpsTilde2D}) becomes divergent, while Eq. (\ref{Sigma-2D}) takes
the already known\cite{Tsuzuki70,Dorsey91,Mishonov02}
form\begin{equation} \sigma
_{\mathrm{NoCut}}^{(2\mathrm{D})}(\widetilde{\varepsilon
},E)=\frac{e^{2}}{16\hbar d}\int _{0}^{\infty }du\,
e^{-\widetilde{\varepsilon }\,
u-\left(\frac{E}{E_{0}}\right)^{2}u^{3}}\,
,\label{Sigma2D-NoCut}\end{equation}
 with the specification that in Eq. (\ref{Sigma2D-NoCut}) the
renormalized
parameter $\widetilde{\varepsilon }$ is present, instead of the reduced
temperature $\varepsilon $.

If, on the contrary, one preserves the cut-off but takes the linear
response limit $E\rightarrow 0$, Eqs. (\ref{Sigma-2D}) and
(\ref{EpsTilde2D}) become\begin{eqnarray}
\left.\sigma ^{(2\mathrm{D})}(\widetilde{\varepsilon })\right|_{E=0} &
= & \frac{e^{2}}{16\hbar d}\left[\frac{1}{\widetilde{\varepsilon
}}-\frac{1}{\widetilde{\varepsilon
}+c}-\frac{c}{(c+\widetilde{\varepsilon })^{2}}\right]\;
,\label{Sigma2D-E0}\\
\widetilde{\varepsilon }-\ln \frac{T}{T_{0}} & = & \frac{2\mu
_{0}\kappa ^{2}e^{2}\xi _{0}^{2}k_{B}T}{\pi \hbar ^{2}d}\ln
\frac{\widetilde{\varepsilon }+c}{\widetilde{\varepsilon }}\;
.\label{EpsTilde2D-E0}
\end{eqnarray}
 A formally identical expression to Eq. (\ref{Sigma2D-E0}) is also
to be found in Ref. \onlinecite{Carballeira01} for Gaussian
fluctuations (i.e. with $\widetilde{\varepsilon }=\varepsilon $), while
Eq. (\ref{EpsTilde2D-E0}) is implicitly contained in the results of
Ref. \onlinecite{Penev0}.

Results for the isotropic three-dimensional (3D) case cannot be
obtained by just imposing the 3D condition $s\rightarrow 0$ (or
$r\rightarrow \infty $) to Eqs. (\ref{Sigma-cutted-off}) and
(\ref{EpsTilde}), because these equations were calculated by assuming
that a cut-off in the $z$ direction is not necessary for layered
superconductors. However, in the 3D case, a cut-off for the
$k_{z}$-momentum is as necessary as for the $k_{x}$ and $k_{y}$
components. The calculations can be performed according to the same
scheme as for the layered case, and the results are presented here for
completeness:\begin{widetext}\begin{eqnarray}
\sigma ^{(3\mathrm{D})}(\widetilde{\varepsilon },E) & = &
\frac{e^{2}}{8\pi \hbar \xi _{0}}\int _{0}^{\infty }du\,
u^{2}e^{-\widetilde{\varepsilon }\,
u-4\left(\frac{E}{E_{0}}\right)^{2}u^{3}}\label{Sigma3D}\\
 &  & \cdot \int _{0}^{c}dw\, w^{3/2}\, e^{-uw}\left[\frac{\cosh
\left(2\sqrt{3}\frac{E}{E_{0}}u^{2}\sqrt{w}\right)}{\left(2\sqrt{3}
\frac{E}{E_{0}}u^{2}\sqrt{w}\right)^{2}}-\frac{\sinh
\left(2\sqrt{3}\frac{E}{E_{0}}u^{2}\sqrt{w}\right)}{\left(2\sqrt{3}
\frac{E}{E_{0}}u^{2}\sqrt{w}\right)^{3}}\right]\, ,\nonumber \\
\widetilde{\varepsilon }-\ln \frac{T}{T_{0}} & = & \frac{2\mu
_{0}\kappa ^{2}e^{2}\xi _{0}k_{B}T}{\pi ^{2}\hbar ^{2}}\int
_{0}^{\infty }du\, e^{-\widetilde{\varepsilon }\,
u-4\left(\frac{E}{E_{0}}\right)^{2}u^{3}}\int _{0}^{c}dw\, w^{1/2}\,
e^{-uw}\, \frac{\sinh
\left(2\sqrt{3}\frac{E}{E_{0}}u^{2}\sqrt{w}\right)}{2\sqrt{3}\frac{E}{E_{0}}u^{2}\sqrt{w}}\,
.\label{EpsTilde3D}
\end{eqnarray}
\end{widetext}

Similarly to the 2D case, if one neglects the cut-off ($c\rightarrow
\infty $), the r.h.s term in Eq. (\ref{EpsTilde3D}) becomes divergent,
while Eq. (\ref{Sigma3D}) takes the expression\begin{equation} \sigma
_{\mathrm{NoCut}}^{(3\mathrm{D})}(\widetilde{\varepsilon
},E)=\frac{e^{2}}{32\sqrt{\pi }\hbar \xi _{0}}\int _{0}^{\infty }du\,
\frac{1}{\sqrt{u}}\, e^{-\widetilde{\varepsilon }\,
u-\left(\frac{E}{E_{0}}\right)^{2}u^{3}}\,
,\label{Sigma3D-NoCut}\end{equation}
 already known\cite{Tsuzuki70,Dorsey91,Mishonov02} for Gaussian
fluctuations
(i.e. with $\widetilde{\varepsilon }=\varepsilon $).

In the linear response limit ($E\rightarrow 0$) but with the cut-off
preserved, Eqs. (\ref{Sigma3D}) and (\ref{EpsTilde3D})
become\begin{eqnarray}
\left.\sigma ^{(3\mathrm{D})}(\widetilde{\varepsilon })\right|_{E=0} &
= & \frac{e^{2}}{48\pi \hbar \xi _{0}}\label{Sigma3D-E0}\\
 & \cdot  & \left[\frac{3\arctan \left(\sqrt{c/\widetilde{\varepsilon
}}\right)}{\sqrt{\widetilde{\varepsilon
}}}-\frac{3\widetilde{\varepsilon }\sqrt{c}}{(\widetilde{\varepsilon
}+c)^{2}}-\frac{5c^{3/2}}{(\widetilde{\varepsilon }+c)^{2}}\right]\,
,\nonumber \\
\widetilde{\varepsilon }-\ln \frac{T}{T_{0}} & = & \frac{4\mu
_{0}\kappa ^{2}e^{2}\xi _{0}k_{B}T}{\pi ^{2}\hbar
^{2}}\left[\sqrt{c}-\sqrt{\widetilde{\varepsilon }}\, \arctan
\left(\sqrt{\frac{c}{\widetilde{\varepsilon }}}\right)\right]\,
.\label{EpsTilde3D-E0}
\end{eqnarray}
 Equation (\ref{Sigma3D-E0}) matches thus formally the expression
found\cite{Carballeira01} for Gaussian fluctuations
($\widetilde{\varepsilon }=\varepsilon $).

\section{Results of the model}

\label{Results}The renormalization procedure required for our present
results consists thus in using the reduced temperature parameter
$\widetilde{\varepsilon }$, renormalized by solving Eq.
(\ref{EpsTilde}), in the conductivity expression
(\ref{Sigma-cutted-off}). This procedure causes the critical
temperature to shift towards lower temperatures. In analogy with the
Gaussian fluctuation case, we shall adopt as definition for the
critical temperature $T_{c}(E)$ the vanishing of the reduced
temperature, $\widetilde{\varepsilon }=0$. Thus, in the absence of the
electric field we can use Eq. (\ref{AnalogUD}) taken at $T=T_{c}(0)$
and $\widetilde{\varepsilon }=0$, so that one gets\begin{equation}
T_{0}=T_{c}(0)\left(\sqrt{\frac{c}{r}}+\sqrt{1+\frac{c}{r}}\right)^{2gT_{c}(0)}\:
.\label{T0}\end{equation}
 In practice, one knows the actual critical temperature $T_{c}(0)\equiv
T_{c0}$
measured at very low electrical field, so that Eq. (\ref{T0}) allows to
estimate the bare mean-field characteristic temperature $T_{0}$. Then,
having the parameter $T_{0}$, one can use Eqs. (\ref{EpsTilde}) for any
temperature $T$ and field $E$ in order to find the actual renormalized
$\widetilde{\varepsilon }(T,E)$, and further the conductivity $\sigma
(T,E)$.

In order to illustrate the main features of the Hartree approximation
for the critical fluctuation model, we shall take as example a common
material, like the optimally doped YBa$_{2}$Cu$_{3}$O$_{6+x}$. Typical
values for the characteristic parameters are then: $s=1.17$ nm for the
interlayer distance, $\xi _{0}=1.2$ nm and $\xi _{0c}=0.14$ nm for the
zero-temperature in-plane and, respectively, out-of-plane coherence
lengths, $\kappa =70$ for the Ginzburg-Landau parameter and $T_{c0}=92$
K for the critical temperature under very small electric field. We also
choose for convenience a linear temperature extrapolation for the
normal state resistivity which vanishes at $T=0$, and has a typical
value $\rho _{N}=84\, \mu \Omega $cm at $T=200$ K.

\begin{figure}
\includegraphics[
width=9cm]{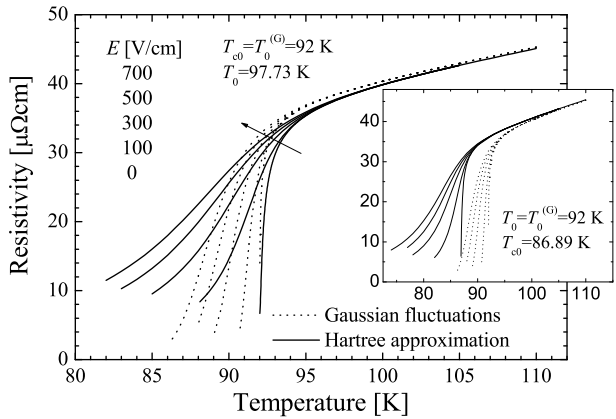} \caption{Resistivity in the Gaussian theory
(dotted curves) and in the Hartree approximation for the interacting
fluctuations (solid curves), for different values of the applied electric
field. The following parameters were used: interlayer distance $s=1.17$
nm; zero-temperature in-plane and out-of-plane coherence lengths, $\xi
_{0}=1.2$ nm and $\xi _{0c}=0.14$ nm, respectively; Ginzburg-Landau
parameter $\kappa =70$; zero-field critical temperature $T_{c0}=92$ K. The
UV cut-off parameter $c=1$ was used. The arrow indicates the increasing
electric field direction. The inset illustrates the critical temperature
shift introduced by the Hartree model if the mean-field transition
temperature $T_{0}$ were kept identical with the one in the Gaussian
approach $T_{0}^{\mathrm{(G)}}$.\label{Resistivity}}
\end{figure}

In Figure \ref{Resistivity} the results of the Hartree approximation
for the critical fluctuations are compared to the ones obtained from
the Gaussian fluctuation theory. The zero-field critical temperature
$T_{c0}$ in the Hartree model was considered identical to the
mean-field critical temperature $T_{0}^{\mathrm{(G)}}$ in the Gaussian
approximation, in order to have the zero-field transition at the same
temperature in both theories. This identification causes the mean-field
transition temperature $T_{0}$ in the Hartree model to shift upwards
with respect to $T_{c0}$. For our chosen parameters this shift was
found to be\begin{equation} T_{0}-T_{c0}=5.733\, \mathrm{K}\:
,\label{T0Shift}\end{equation}
 while taking a cut-off parameter $c=1$. The difference between the
two temperatures depends on choice of the cut-off parameter, namely it
increases with the $c$ value, and becomes divergent for no cut-off
($T_{0}/T_{c0}\rightarrow \infty $ for $c\rightarrow \infty $). It can
be noticed in Figure \ref{Resistivity} that the curves obtained in the
Hartree approximation are less steep than those in the Gaussian one,
and show a significantly broadened transition region in the presence of
strong applied electric fields. In addition, we find that the
paraconductivity in the renormalized model is more sensitive to the
electric field, showing a more pronounced suppression of the
fluctuations at high fields in the lower part of the transition.
However, above the zero-field transition temperature, the
paraconductivity in the Hartree model is always higher than the one in
the Gaussian approximation, due essentially to the critical temperature
redefinition from $T_{0}$ to $T_{c0}$. If one preserved instead the
same mean-field transition temperature $T_{0}=T_{0}^{\mathrm{(G)}}$ as
in the Gaussian approximation, one could then visualize the critical
temperature shift introduced by the Hartree approach, as shown in the
inset of Fig. \ref{Resistivity}. Equation (\ref{T0}) would give then
$T_{c0}=86.894\, $K, and the paraconductivity would be always lower
than the one in the Gaussian approximation.

\begin{figure}
\includegraphics[  width=8cm]{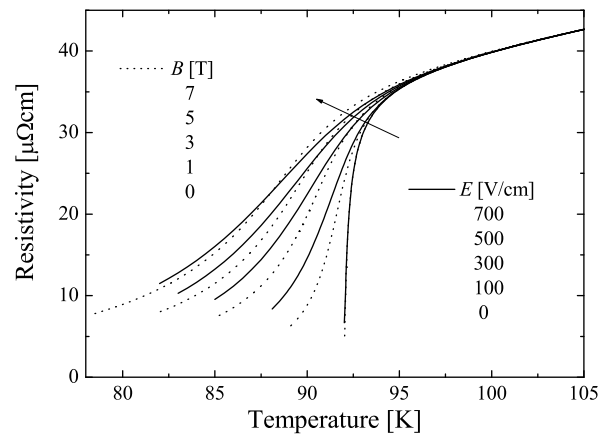}
\caption{Comparison between the transition broadening effect of the
electric (solid lines) and magnetic (dotted lines) field, respectively,
according to the renormalized Hartree approximation. The same
parameters as in Fig. \ref{Resistivity} were used. The magnetic field
effect was calculated in the linear response (zero electric field)
limit, according to the UD model.\cite{Ullah91} The UV cut-off
parameter $c=1$ in our model corresponds with limiting the sum on the
Landau levels at the index $1/2h$ in the UD model ($h=2\pi \xi
_{0}^{2}B/\Phi _{0}$ being the reduced magnetic field). The arrow
indicates the increasing electric and magnetic field
direction.\label{ComparisonBE}}
\end{figure}

For illustration, we also give in Figure \ref{ComparisonBE} a
comparison between the results of our model, applicable for layered
superconductors in arbitrary electric fields in the absence of magnetic
field, and the ones of the complementary model of Ullah and
Dorsey,\cite{Ullah91} which treats in the same Hartree approximation
the case of an arbitrary magnetic field in the linear response (zero
electric field) limit. As one can easily notice, the same well known
{}``fan shape'' transition broadening, encountered when a magnetic
field is applied, can be also predicted for the presence of a
sufficiently strong electric field. We argue thus that high electric
fields can be used to suppress order parameter fluctuations in HTSC as
effectively as a magnetic field.

This comparison between the fluctuation suppression effects of the
electric and magnetic field could give also a rough estimation on how
broad the validity domain of our model could be. It is known for
instance that the renormalized fluctuation model\cite{Ikeda91a} for
layered superconductors can successfully fit the resistivity curves of
YBCO and BSCCO single crystals in magnetic fields up to about 10 T, in
a temperature range that covers approximately the two superior thirds
of the transition region. This extends from a few K in lower fields, up
to more than 10 K in higher fields, measured down from the resistivity
onset point in zero field.\cite{Ikeda91a} The lowest third of the
transition, where, experimentally, the resistivity slope becomes
steeper, is instead affected by flux pinning effects and does not fit
into the GL theory. We can therefore assume, based on the similarities
illustrated in Fig. \ref{ComparisonBE}, that also in high electric
fields, the renormalized fluctuation model based on the TDGL approach
may have its validity in a temperature range at least as broad as in
the case when a magnetic field is applied. Since under high electric
fields (and consequently, high current densities), the pinning of the
self-field flux lines is overcome by the high Lorentz force, we can
expect that the validity of the presented model might extend for even
lower temperatures.

In a few previous papers,\cite{Soret93,Gorlova95} the non-ohmic
fluctuation conductivity in high electric fields was reported to be
experimentally proven, by confronting the measured paraconductivity to
the scaling laws predicted by Gaussian fluctuation
models.\cite{Schmid69,Varlamov92} However, the broadening of the
temperature dependence of the resistive transition with respect to the
increasing electric field, and the breaking of the mean-field
(Gaussian) theory in the immediate vicinity of $T_{c}$, signalized by
Ref. \onlinecite{Soret93}, indicate that a renormalized (non-Gaussian)
fluctuation model, as the one presented in this paper, might be more
appropriate. From the experimental viewpoint, applying electric fields
of a few hundreds V/cm on cuprates may however be not an easy task,
since the dissipated power density would attain levels of the order of
$\mathrm{GWcm}^{-3}$. On the one hand, high electric fields are
necessary in order to put into evidence the non-ohmic fluctuation
conductivity, while, on the other hand, they produce high dissipation
and can increase the sample temperature at values where the
nonlinearity is no longer discernable. In this connection, using short
current pulses at high current densities (a few $\mathrm{MAcm}^{-2}$),
seems to be a better alternative to the dc and ac measurements.

\section{Conclusions}

\label{Conclusion}In summary, we have treated in this paper the
critical fluctuation conductivity for a layered superconductor in zero
magnetic field, in the frame of the self-consistent Hartree
approximation, for an arbitrary electric field magnitude. The main
results of our work are the formulae (\ref{Sigma-cutted-off}) for the
fluctuation conductivity, and (\ref{EpsTilde}) for the renormalized
reduced-temperature parameter. In both equations the UV cut-off of the
Ginzburg-Landau model was explicitly considered. In the linear-response
limit ($E\rightarrow 0$), the corresponding expressions Eqs.
(\ref{LD-cutOff}) and (\ref{AnalogUD}) reduce to the previous results
of existing theories. Qualitatively, the temperature characteristics at
different electrical fields in the Hartree approximation turn out to be
less steep than those in the Gaussian one, they show a more pronounced
suppression of the fluctuations at high fields in the lower part of the
transition, and a higher paraconductivity above the zero-field
transition temperature than the Gaussian fluctuation model. All these
features are quantitatively important for commonly used HTSC, so that
experimental investigations could be able to discern easily between the
applicability of this model in competition with the Gaussian
fluctuation approximation.

\begin{acknowledgments}
This work was supported by the Austrian Fonds zur F\"{o}rderung der
wissenschaftlichen Forschung. Stimulating correspondence and
discussions with R. Ikeda and A.A. Varlamov are also gratefully
acknowledged.
\end{acknowledgments}
\appendix*

\section{Green function for the TDGL equation}

\label{App1}Equation (\ref{GreenEq}) can be solved easier for the
Fourier transform of the Green function with respect to time,
\begin{equation} R_{q}(\mathbf{k},\omega ;k'_{x},t')=\int dt\;
R_{q}(\mathbf{k,}t;k'_{x},t')\; e^{i\omega (t-t')}\;
,\label{TimeFourier}\end{equation}
 which satisfies the equation\begin{widetext} \begin{eqnarray}
\left[-i\omega \Gamma _{0}^{-1}-\frac{e_{0}\Gamma _{0}^{-1}E}{\hbar
}\frac{\partial }{\partial _{k_{x}}}+\frac{\hbar
^{2}k_{x}^{2}}{2m}+a_{1}\right]R_{q}(\mathbf{k},\omega
;k'_{x},t')=\delta (k_{x}-k'_{x})\; . &  & \label{EQtimeFour}
\end{eqnarray}
 One finds for the differential Eq. (\ref{EQtimeFour}) the solution
\begin{eqnarray}
R_{q}(\mathbf{k},\omega ;k'_{x},t') & = & A_{q}(\mathbf{k},\omega
;k'_{x},t')\cdot \exp \left\{ \frac{\hbar \Gamma
_{0}}{e_{0}E}\left[\frac{\hbar ^{2}k_{x}^{3}}{6m}+\left(a_{1}-i\omega
\Gamma _{0}^{-1}\right)k_{x}\right]\right\} \; ,\label{Homogeneous}
\end{eqnarray}
 where the derivative of the coefficient $A_{q}(\mathbf{k},\omega
;k'_{x},t')$
must satisfy\begin{eqnarray}
\frac{\partial A_{q}}{\partial k_{x}}(\mathbf{k},\omega ;k'_{x},t') & =
& -\frac{\hbar \Gamma _{0}}{e_{0}E}\exp \left\{ -\frac{\hbar \Gamma
_{0}}{e_{0}E}\left[\frac{\hbar ^{2}k_{x}^{3}}{6m}+\left(a_{1}-i\omega
\Gamma _{0}^{-1}\right)k_{x}\right]\right\} \delta
(k_{x}-k'_{x})\nonumber \\
 & = & -\frac{\hbar \Gamma _{0}}{e_{0}E}\exp \left\{ -\frac{\hbar
\Gamma _{0}}{e_{0}E}\left[\frac{\hbar ^{2}k_{x}^{\prime
3}}{6m}+\left(a_{1}-i\omega \Gamma
_{0}^{-1}\right)k'_{x}\right]\right\} \delta
(k_{x}-k'_{x}).\label{CoeffDeriv}
\end{eqnarray}
 The solution for the coefficient $A_{q}(\mathbf{k},\omega
;k'_{x},t')$,
which remains nondivergent when $k'_{x}\rightarrow -\infty $, is
\begin{equation}
A_{q}(\mathbf{k},\omega ;k'_{x})=\frac{\hbar \Gamma _{0}}{e_{0}E}\theta
(k'_{x}-k_{x})\exp \left\{ -\frac{\hbar \Gamma
_{0}}{e_{0}E}\left[\frac{\hbar k_{x}^{\prime
3}}{6m}+\left(a_{1}-i\omega \Gamma
_{0}^{-1}\right)k'_{x}\right]\right\} \label{Coeff}\end{equation}
 where $\theta (k'_{x}-k_{x})$ is the Heavyside step function, so
that the Fourier transform of the Green function will be
\begin{equation} R_{q}(\mathbf{k},\omega ;k'_{x},t')=\frac{\hbar \Gamma
_{0}}{e_{0}E}\; \theta (k'_{x}-k_{x})\cdot \exp \left\{ \frac{\hbar
\Gamma _{0}}{e_{0}E}\left[\frac{\hbar ^{2}\left(k_{x}^{3}-k_{x}^{\prime
3}\right)}{6m}+\left(a_{1}-i\omega \Gamma
_{0}^{-1}\right)\left(k_{x}-k'_{x}\right)\right]\right\}
\label{FourierGreenSol}\end{equation}
 Now we can apply the inverse Fourier transform to regain the Green
function depending on time $t$, and obtain \begin{eqnarray}
 &  & R_{q}(\mathbf{k,}t;k'_{x},t')=\int \frac{d\omega }{2\pi }\;
R_{q}(\mathbf{k},\omega ;k'_{x},t')\; e^{-i\omega
(t-t')}\label{GreenSol}\\
 &  & =\frac{\hbar \Gamma _{0}}{e_{0}E}\; \theta (k'_{x}-k_{x})\exp
\left\{ \frac{\hbar \Gamma _{0}}{e_{0}E}\left[\frac{\hbar
^{2}\left(k_{x}^{3}-k_{x}^{\prime
3}\right)}{6m}+a_{1}\left(k_{x}-k'_{x}\right)\right]\right\} \nonumber
\\
 &  & \cdot \int \frac{d\omega }{2\pi }\exp \left\{ -i\omega
\left[t-t'+\frac{\hbar
}{e_{0}E}\left(k_{x}-k'_{x}\right)\right]\right\} \nonumber \\
 &  & =\frac{\hbar \Gamma _{0}}{e_{0}E}\; \theta (k'_{x}-k_{x})\exp
\left\{ \frac{\hbar \Gamma _{0}}{e_{0}E}\left[\frac{\hbar
^{2}\left(k_{x}^{3}-k_{x}^{\prime
3}\right)}{6m}+a_{1}\left(k_{x}-k'_{x}\right)\right]\right\} \delta
\left(t-t'+\frac{\hbar }{e_{0}E}\left[k_{x}-k'_{x}\right]\right)\,
.\nonumber
\end{eqnarray}
 \end{widetext}The form (\ref{Coeff}) for the coefficient
$A_{q}(\mathbf{k},\omega ;k'_{x})$,
and namely the presence of the Heavyside function $\theta
(k'_{x}-k_{x})$ assures that the Green function
$R_{q}(\mathbf{k},\omega ;k'_{x},t')$ in Eq. (\ref{FourierGreenSol})
doesn't diverge, and provides also the retarded character in Eq.
(\ref{GreenSol}), i.e. $R_{q}(\mathbf{k,}t;k'_{x},t')=0$ for $t<t'$.

\bibliographystyle{APSREV}
\bibliography{puica_lang}

\end{document}